\input{aipcheck}

\documentclass[
    ,final            
  ]
  {aipproc}

\layoutstyle{6x9}

\begin{document}

\title{New generation of double beta decay experiments: are there any limitations?}

\classification{23.40-s, 14.60.Pq}
\keywords      {double beta decay, half-life values}

\author{A.S. Barabash}{
  address={Institute of Theoretical and Experimental Physics, B.\
Cheremushkinskaya 25, 117259 Moscow, Russia}
}

\begin{abstract}
New generation experiments on search for neutrinoless double beta
decay with sensitivity to effective Majorana neutrino mass on the level $\sim$ 3-5 meV is discussed. 
Possible restrictions at achievement of this purpose (possibility to produce big amount of enriched 
isotopes, possibility to reach very low background level, energy resolution and 
possible cost of experiments) are considered. It is shown that for realization of so ambitious 
project 10 tons (or more) of enriched isotope 
is required. Background index should be on the level $\le$ 10$^{-5}$-10$^{-6}$ c/kg$\cdot$ keV$\cdot$ y. 
Besides, the energy 
resolution of the detector should be not worse than 1-2\%. It is shown that $^{130}$TeO$_2$ low temperature 
bolometer looks as the most realistic candidate for such experiments. 
Under some conditions experiments with $^{76}$Ge, $^{100}$Mo and $^{136}$Xe can be realized too.

\end{abstract}

\maketitle

\section{Introduction}

 The $0\nu\beta\beta$-decay rate depends on the type of neutrino mass spectrum which can 
be hierarchical, with partial hierarchy or quasi-degenerate (see, e.g., \cite{BIL01}). 
Using the data on the neutrino oscillation parameters it is possible to show 
(see, e.g., \cite{NAK10}) that in the case of normal hierarchical spectrum one has 
|$\langle m_{\nu }\rangle $| < 0.005eV, while if the spectrum is with inverted hierarchy, 
0.01 eV < |$\langle m_{\nu }\rangle $| < 0.05 eV. A larger value of |$\langle m_{\nu }\rangle $| is possible if the light 
neutrino mass spectrum is with partial hierarchy or of quasi-degenerate type. 
In the latter case |$\langle m_{\nu }\rangle $| can be close to the existing upper limits. 
In the present work a current state of experiments on search for double beta 
decay (with sensitivity to |$\langle m_{\nu }\rangle $| $\sim$ 0.2-1 eV) is analysed, offers on immediate 
prospects (|$\langle m_{\nu }\rangle $| $\sim$ 10-100 meV) is considered and possibility to achieve 
the sensitivity $\sim$ 1-5 meV is estimated.

\section{Present status}

The constraints on the existence of $0\nu\beta\beta$-decay are presented in Table 1 for the nuclei for
which the best sensitivity has been reached. In calculating constraints on |$\langle m_{\nu }\rangle $|, the nuclear
matrix elements (NMEs) from \cite{KOR07,KOR07a,SIM08,BAR09,RAT10,ROD10} were used (3-d column). 
In column four, limits on |$\langle m_{\nu }\rangle $|, which were 
obtained using the NMEs from a recent Shell Model (SM) calculations \cite{CAU08} are presented (for 
$^{116}$Cd NME from \cite{CAU07} is used). And now the limits on |$\langle m_{\nu }\rangle $|  
for $^{130}$Te and $^{100}$Mo are comparable with the $^{76}$Ge results. The assemblage of sensitive experiments 
for different nuclei permits one to increase the reliability of the limit on |$\langle m_{\nu }\rangle $|. Present 
conservative limit can be set as 0.75 eV.
     In Table 2 the reached level of background in the mentioned above experiments is presented. 
Background index (BI) and sum background ($\sum$B) in region of interest (ROI) are given. One can see that BI 
in present experiments is quite high and one has a deal with background in ROI (there is no any 
"zero" background experiment up to now).

\begin{table}[h]
\caption{Best present results on $2\beta(0\nu)$ decay (limits at
90\% C.L.). $^{*)}$ See discussions in \cite{BAR10}; $^{**)}$ NME from \cite{CAU07} is used;
$^{***)}$ conservative limit from \cite{BER02} is presented}
\begin{tabular}{|c|c|c|c|c|}
\hline
Isotope & $T_{1/2}$, y & |$\langle m_{\nu} \rangle$|, eV & |$\langle
m_{\nu} \rangle$|, eV & Experiment \\
& & \cite{KOR07,KOR07a,SIM08,BAR09,RAT10,ROD10} & \cite{CAU08}  \\
\hline
$^{76}$Ge & $>1.9\cdot10^{25}$ & $<0.22-0.41$ & $<0.69$ & HM
\cite{KLA01} \\
& $\simeq 1.2\cdot10^{25}$(?)$^{*)}$ & $\simeq 0.28-0.52(?)$$^{*)}$ & $\simeq
0.87(?)$$^{*)}$ & Part of HM \cite{KLA04} \\
& $\simeq 2.2\cdot10^{25}$(?)$^{*)}$ & $\simeq 0.21-0.38(?)$$^{*)}$ & $\simeq
0.64(?)$$^{*)}$ & Part of HM \cite{KLA06} \\
& $>1.6\cdot10^{25}$ & $<0.24-0.44$ & $<0.75$ & IGEX
\cite{AAL02} \\
\hline
$^{130}$Te & $>2.8\cdot10^{24}$ & $<0.29-0.59$ & $<0.77$ &
CUORICINO \cite{AND11} \\
$^{100}$Mo & $>1.1\cdot10^{24}$ & $<0.29-0.93$ & $ - $ & NEMO-
3 \cite{BAR11} \\
$^{136}$Xe & $>4.5\cdot10^{23}$$^{***)}$ & $<1.41-2.67$ & $<2.2$ & DAMA
\cite{BER02} \\
$^{82}$Se & $>3.6\cdot10^{23}$ & $<1.89-1.61$ & $<2.3$ & NEMO-3
\cite{BAR11} \\
$^{116}$Cd & $>1.7\cdot10^{23}$ & $<1.45-2.76$ & $<1.8$$^{**)}$ &
SOLOTVINO \cite{DAN03} \\
\hline
\end{tabular}
\end{table}

\begin{table}[h]
\caption{Background index (BI) and sum background in ROI ($\sum$B) in the best present experiments. 
M- mass of investigated isotope, t - measurement time, Q - energy of 2$\beta$-decay, 
$\Delta$E/E - energy resolution (FWHM). $^{*)}$ After pulse shape analysis; $^{**)}$ for $^{130}$Te.}
\begin{tabular}{|c|c|c|c|c|}
\hline 
Experiment & M$\cdot$t, kg$\cdot$ y & $\Delta$E/E, \% at Q & BI, c/kg$\cdot$ keV$\cdot$ y & $\sum$B (ROI=$\Delta$E) \\
\hline
HM & 71 &  0.2 & 0.17 (0.02)$^{*)}$ & $\sim$ 50 ($\sim$ 3)$^{*)}$ \\
IGEX & 7 & 0.2 & 0.2 (0.06)$^{*)}$ & $\sim$ 7 ($\sim$ 2)$^{*)}$ \\
CUORICINO & 72 (20)$^{**)}$ & 0.3 & 0.18 & $\sim$ 70 \\
NEMO-3 ($^{100}$Mo) & 31  & 8 & 1.4$\cdot$ 10$^{-3}$ & $\sim$ 18 \\
DAMA & 6.5 & 20 & 0.08 & $\sim$ 250 \\
SOLOTVINO & 0.53 & 9 & 0.04 & $\sim$ 5 \\
\hline
\end{tabular}
\end{table}

\section{Next generation of double beta decay experiments (|$\langle m_{\nu }\rangle $| $\sim$ 10-100 meV)}

There are a few tens of different propositions for future double beta decay experiments. Here
seven of the most developed and promising experiments which can be realized within the next
few years are presented (see Table 3). The estimation of the sensitivity in the experiments is
made using NMEs from \cite{KOR07,KOR07a,SIM08,BAR09,RAT10,ROD10,CAU08}. In all probability, they will make it possible to
reach the sensitivity for the neutrino mass at a level of 0.01 to 0.1 eV.
     In Table 4 planned BI and $\sum$B in ROI are presented. For BI we are waiting for 
100-1000 times lower values in comparison with present experiments. And even under 
such improvements $\sum$B will be non-zero. But in some cases (MAJORANA, EXO, SuperNEMO) it will 
be just a few events after 5-10 years of measurements. 

\begin{table}[h]
\caption{Seven most developed and promising projects. 
Sensitivity at 90\% C.L. for three (1-st step of GERDA and MAJORANA, SNO+, and KamLAND-Xe) 
five (EXO, SuperNEMO and CUORE) and ten (full-scale GERDA and MAJORANA) 
years of measurements is presented. M - mass of isotopes.}
\begin{tabular}{c|c|c|c|c|c}
\hline
Experiment & Isotope & M, kg & Sensitivity & Sensitivity & Status \\
& &  & $T_{1/2}$, y & |$\langle m_{\nu} \rangle$|, meV &  \\
\hline
CUORE \cite{ARNA04} & $^{130}$Te & 200 & $2.1\times10^{26}$ & 35--90 & in progress \\ 
GERDA \cite{ABT04} & $^{76}$Ge & 40 & $2\times10^{26}$ & 70--300 & in progress \\
& & 1000 & $6\times10^{27}$ & 10--40 & R\&D\\ 
MAJORANA & $^{76}$Ge & 30--60 & (1--2)$\times10^{26}$ & 70--300 & in progress \\
\cite{MAJ03, AVI10}& & 1000 & $6\times10^{27}$ & 10--40 & R\&D \\ 
EXO \cite{DAN00} & $^{136}$Xe & 200 & $6.4\times10^{25}$ & 95--220 & in progress \\
& & 1000 & $8\times10^{26}$ & 27--63 & R\&D \\ 
SuperNEMO & $^{82}$Se & 100--200 & (1--2)$\times10^{26}$ & 40--110 & R\&D \\
\cite{BAR02,BAR04a,SAA09} & & & & &\\
KamLAND-Xe & $^{136}$Xe & 400 & 4.5$\times10^{26}$ & 40--80 & in progress \\
\cite{NAK10a} & & 1000 &  $\sim$ 10$^{27}$ & 25-50 & R\&D \\
SNO+ \cite{KRA10} & $^{150}$Nd & 56 & $\sim$ 4.5$\times10^{24}$ & 100--300 & in progress \\
 & & 500 & $\sim$ 3$\times10^{25}$ & 40-120 &  R\&D \\
\hline
\end{tabular}
\end{table}

\begin{table}[h]
\caption{Background index (BI) and sum background in ROI ($\sum$B) in the next generation experiments. 
M-mass of isotope, Q - energy of 2$\beta$-decay. $\sum$B is given for the measurement time indicated in Table 3.
$^{*)}$ Full weight of the detector.}
\begin{tabular}{c|c|c|c|c|c}
\hline
Experiment & Isotope & M, kg & $\Delta$E/E at Q, \% & BI, c/kg$\cdot$ keV$\cdot$ y & $\sum$B (ROI=$\Delta$E) \\
\hline
CUORE & $^{130}$Te & 200 (740)$^{*)}$ & 0.3 & 0.01 & $\sim$ 180 \\ 
GERDA & $^{76}$Ge & 40 & 0.16 & 0.001 & $\sim$ 0.4 \\
& & 1000 & 0.16 & < 0.001 & < 30\\ 
MAJORANA & $^{76}$Ge & 30--60 & 0.16 & 0.001 & $\sim$ 0.3-0.6 \\
& & 1000 & 0.16 & 0.00025 & $\sim$ 8 \\ 
EXO & $^{136}$Xe & 200 & 3.8 & 0.001 & $\sim$ 100 \\
& & 1000 & 3.8 & $\sim$ 2$\cdot$ 10$^{-6}$ & $\sim$ 1 \\ 
SuperNEMO & $^{82}$Se & 100--200 & 4-5 & $\sim$ 2$\cdot$ 10$^{-5}$ & $\sim$ 1-2 \\
KamLAND-Xe & $^{136}$Xe & 400 ($1.6\cdot10^4$)$^{*)}$ & 10 & $\sim$ 10$^{-6}$ & $\sim$ 15 \\
& & 1000 ($4\cdot10^4$)$^{*)}$ & & &$\sim$ 40 \\
SNO+ & $^{150}$Nd & 56 ($10^6$)$^{*)}$ & 6.4 & $\sim$ 10$^{-6}$ & $\sim$ 600 \\
 & & 500 ($10^6$)$^{*)}$ &  &  & $\sim$ 600 \\
\hline
\end{tabular}
\end{table}

\section{New generation of double beta decay experiments (|$\langle m_{\nu }\rangle $| $\sim$  1-5 meV)}

Table 5 presents number of nuclei in 10 tons of different isotopes and number of events obtained 
with 10 t of isotope after 10 y of measurement and for $T_{1/2} = 10^{29}$ y. In Table 6 estimated half-life 
values for different isotopes and for |$\langle m_{\nu }\rangle $| = 1, 3 and 5 meV are presented. 
In bold $T_{1/2}$ values at which disintegration can be registered are allocated.
Using information from Tables 5 and 6 one can conclude that with 10 t of isotope sensitivity to
|$\langle m_{\nu }\rangle $| on the level 3-5 meV can be reached with some isotopes. The best sensitivity 
can be reached with $^{100}$Mo and $^{150}$Nd. With $^{136}$Xe it will be difficult to reach even 5 meV sensitivity.
And for $^{48}$Ca the best possible sensitivity is estimated as $\sim$ 7 meV only.

\begin{table}[h]
\caption{Number of nuclei in 10 t of isotope and number of events after 10 years of measurement 
(for $T_{1/2} = 10^{29}$ y).}
\begin{tabular}{|c|c|c|}
\hline
Isotope & N of nuclei in 10 t of isotope &  Events per 10 t and 10 y ($T_{1/2} = 10^{29}$ y) \\
\hline
$^{48}$Ca & 1.25$\cdot$ 10$^{29}$ &  8.6   \\
$^{76}$Ge & 7.9$\cdot$ 10$^{28}$ & 5.5  \\
$^{82}$Se & 7.3$\cdot$ 10$^{28}$  & 5  \\
$^{100}$Mo & 6$\cdot$ 10$^{28}$  & 4.1 \\
$^{116}$Cd & 5.2$\cdot$ 10$^{28}$  & 3.6  \\
$^{130}$Te & 4.6$\cdot$ 10$^{28}$  & 3.2  \\
$^{136}$Xe & 4.4$\cdot$ 10$^{28}$  & 3  \\
$^{150}$Nd & 4$\cdot$ 10$^{28}$  & 2.8  \\

\hline
\end{tabular}
\end{table}

\begin{table}[h]
\caption{Half-life values (in yr) for different values of |$\langle m_{\nu }\rangle $|. NME values from 
\cite{KOR07,KOR07a,SIM08,BAR09,RAT10,ROD10,CAU08} were used.}
\begin{tabular}{|c|c|c|c|}
\hline
Isotope & |$\langle m_{\nu }\rangle $| = 1 meV & |$\langle m_{\nu }\rangle $| = 3 meV & |$\langle m_{\nu }\rangle $| = 5 meV \\
\hline
$^{48}$Ca & 1.1$\cdot$ 10$^{31}$ & 1.2$\cdot$ 10$^{30}$  & 4.4$\cdot$ 10$^{29}$ \\
$^{76}$Ge & (0.9-9)$\cdot$ 10$^{30}$ & ({\bf 0.1}-1)$\cdot$ 10$^{30}$ & ({\bf 0.37}-3.6)$\cdot$ 10$^{29}$  \\
$^{82}$Se & (0.28-1.9)$\cdot$ 10$^{30}$ & ({\bf 0.3}-2.1)$\cdot$ 10$^{29}$ & {\bf (1.1-7.6)$\cdot$ 10$^{\bf 28}$}  \\
$^{100}$Mo & ({\bf 0.9}-9.4)$\cdot$ 10$^{29}$ & {\bf (0.1-1)$\cdot$ 10$^{\bf 29}$} & {\bf (0.37-3.8)$\cdot$ 10$^{\bf 28}$}  \\
$^{116}$Cd & (0.36-1.3)$\cdot$ 10$^{30}$ & ({\bf 0.4}-1.4)$\cdot$ 10$^{29}$ & {\bf (1.4-5.2)$\cdot$ 10$^{\bf 28}$}  \\
$^{130}$Te & (0.24-1.7)$\cdot$ 10$^{30}$ & ({\bf 0.27}-1.9)$\cdot$ 10$^{29}$ & {\bf (1-6.8)$\cdot$ 10$^{\bf 28}$}  \\
$^{136}$Xe & (0.89-3.2)$\cdot$ 10$^{30}$ & (1-3.6)$\cdot$ 10$^{29}$ & ({\bf 0.36}-1.4)$\cdot$ 10$^{29}$  \\
$^{150}$Nd & (1.2-4.2)$\cdot$ 10$^{29}$ & {\bf (1.3-4.7)$\cdot$ 10$^{\bf 28}$} & {\bf (0.48-1.7)$\cdot$ 10$^{\bf 28}$}  \\
\hline
\end{tabular}
\end{table}

\subsection{Possible experimental approaches}   

Let's consider possible experimental approaches to such measurements:

- HPGe detectors;

- low temperature bolometers;

- liquid scintillator detectors (KamLAND, SNO+, SK+, BOREXINO);

- liquid (or gas) Xe detectors (EXO, XMASS, NEXT);

- new ideas - !?

Most of these approaches are used in present experiments (see reviews \cite{BAR10,BAR11a}). And I hope that the new ideas are coming.

\subsection{Possible background limitations}

Background conditions are the key point for $2\beta$-decay experiments. To detect 
the $0\nu\beta\beta$-decay one has to detect  (as minimum) $\sim$ 5-10 events and 
background has to be $\sim$ 0-2 events only! Say, for HPGe detectors BI has 
to be < $5\cdot$ 10$^{-6}$ c/kg$\cdot$ keV$\cdot$ y ($\sim$ 100 times better then planed background 
in MAJORANA). Background will be a real problem for next generation 
experiments. Main sources of background are the following: 

- contaminations in detector and shield;

- cosmic rays;

- 2$\nu$ tail;

- solar, reactor and geo neutrinos.

       Of course, needed purity is differ for different experiments. 
But, in any case, it is better to have "clever" detector, which can 
recognize $2\beta$ events (granularity, anticoincidence, tracks reconstruction, 
daughter ions registration and so on). It is well known that in BOREXINO, SNO, 
KamLAND purity of different liquids and gazes is on the level 
$\sim$ 10$^{-16}$-10$^{-17}$ g/g of U and Th. In principle, solid material can be purified 
to the same level (in present experiments it is $\sim$ 10$^{-12}$ g/g). So, 
in principle, one can have pure enough materials for 10 t 2$\beta$-decay experiments. 
But it will take a lot of efforts, time and money.
     Concerning to cosmic rays, main background is connected with muons 
itself, $\gamma$ and neutrons induced by cosmic ray muons and radioactive isotopes produced by muons. 
Main recipe here is to go deep underground (6000 m w.e. or more)  
and to use effective veto shield. In Ref. \cite{MEI06} it was demonstrated that 
 BI = 10$^{-6}$ c/keV$\cdot$ kg$\cdot$ y can be obtained for HPGe detectors.
     To avoid contribution from 2$\nu$ tail energy resolution has to be 
better then 1-2\% (see discussion in \cite{ZDE04}).
     Recently it was demonstrated that BI connected with solar neutrinos will be on the level 
$\sim$ (1-2)x10$^{-7}$ c/keV$\cdot$ kg$\cdot$ y \cite{ZUB11}. So, if energy resolution is good enough (say, 1-2\%) this 
contribution will be negligible in most cases. Background from reactor and 
geo neutrino are in $\sim$ 10 and $\sim$ 100 times lower \cite{KLI04}.

\subsection{Possibilities of double beta isotope production}
       There are different methods for isotope 
production:

- centrifugation (productivity in arbitrary units is 1);

- laser separation (productivity is $\sim$ 0.1);

- plasma separation (productivity is $\sim$ 0.01);

- electromagnetic separation (productivity is $\sim$ 0.001).

Taking into account the productivity and cost (which is proportional to productivity)
it is clear that centrifugation is the only method to produce 10 tons of 
enriched material for 2$\beta$-decay experiments. Present productivity one can estimate as 
$\sim$ 200 kg per year. It can be increased in $\sim$ 10 times (with 
additional money investment). So, 10 t can be produced during 5-10 years. 
If it will be necessary new facility can be organized for this goal. 

 \begin{table}[h]
\caption{Approximate price of 2$\beta$ isotopes obtained by senrifugation. 
$^{*)}$ Taking into account 20\% reduction for mass production case.}
\begin{tabular}{|c|c|c|c|}
\hline
Isotope & Abundance & Price per kg, kS &  Cost of 10 t, Mln.S\\
\hline

$^{76}$Ge & 7.61 & $\sim$ 80 & 800  (640)$^{*)}$  \\
$^{82}$Se & 8.73 & $\sim$ 120 & 1200 (1000)$^{*)}$  \\
$^{100}$Mo & 9.63 & $\sim$ 80 & 800 (640)$^{*)}$  \\
$^{116}$Cd & 7.49 & $\sim$ 180 & 1800 (1440)$^{*)}$  \\
$^{130}$Te & 34.08 & $\sim$ 20 & 200 (160)$^{*)}$  \\
$^{136}$Xe & 8.87 & $\sim$ 5-10 & 50-100 (40-80)$^{*)}$  \\
$^{150}$Nd (?) & 5.6 & > 200 & > 2000  \\
\hline
\end{tabular}
\end{table}

\subsection{Cost of experiments} 
Some cost estimations are presented in Table 7. One can see that cheapest 
possibilities are $^{136}$Xe and $^{130}$Te. $^{76}$Ge and $^{100}$Mo are on the border of 
money possibilities. For other isotopes cost start to be real limitation. 
     In case of $^{136}$Xe there is another broblem. Xenon is very rear material, 
its concentration in atmosphere is $\sim$ 10$^{-5}$\%. World rate production is $\sim$ 40 tons 
per year. To collect 10 t of $^{136}$Xe one will need 100 t of natural Xe. It means 
that it will be very difficult (if possible) to have 10 t of $^{136}$Xe. But to reach 3 meV sensitivity 
region one will need $\sim$ 20-30 t of $^{136}$Xe (see Table 6).

\section{Conclusion}

Taking into account all above arguments one can conclude; 

1) 10 t detector with sensitivity to neutrino mass on the level $\sim$ 3-5 meV can be created
using existing techniques.

2) Strong international collaboration will be needed.

3) Minimal cost of such experiment is $\sim$ 100-300 Mln. dollars.

4) $^{130}$TeO$_2$ low temperature bolometer looks as the best candidate for such experiments.
In this case even natural Te can be used.

5) $^{130}$Xe is good candidate too (EXO or NEXT type detectors) if it will be possible to produce 20-30 t of enriched Xe.

6) HPGe detector made of enriched Ge and low temperature bolometer containing $^{100}$Mo 
could be used too if it will be possible to decrease cost of enriched Ge and Mo
production.

In any case, we have to wait, first, for results with CUORE, MAJORANA/GERDA, EXO and other experiments to be sure that 
all mentioned above problems can be solved.
And, of course, new ideas are needed.






\bibliographystyle{aipproc}   

\bibliography{sample}

\begin{thebibliography}{99}
\bibitem{BIL01}
S.M. Bilenky, S. Pascoli and S.T. Petcov, {\it Phys. Rev. D} {\bf 64} (2001) 053010 and 113003.
\bibitem{NAK10}
K. Nakamura et al. (Particle Data Group) {\it J. Phys. G} {\bf 37} (2010) 075021.
\bibitem{KLA01}
H. V. Klapdor-Kleingrothaus et al., {\it Eur. Phys. J. A} {\bf 12} (2001) 147.
\bibitem{KLA04}
H. V. Klapdor-Kleingrothaus et al., {it Phys. Lett. B} {\bf 586} (2004) 198.
\bibitem{KLA06}
H .V. Klapdor-Kleingrothaus et al., {\it Mod. Phys. Lett. A} {\bf 21} (2006) 1547.
\bibitem{BAR10}
A.S. Barabash, {\it Phys. At. Nucl.} {\bf 73} (2010) 162.
\bibitem{AAL02}
C. E. Aalseth et al., {it Phys. Rev. D} {\bf 65} (2002) 092007.
\bibitem{AND11}
E. Andreotti et al., {it Astropart. Phys.} {\bf 34} (2011) 822.
\bibitem{BAR11}
A.S. Barabash and V.B. Brudanin, {\it Phys. At. Nucl} {\bf 74} (2011) 312.
\bibitem{BER02}
R. Bernabei et al., {it Phys. Lett. B} {\bf 546} (2002) 23.
\bibitem{DAN03}
F. A. Danevich et al., {it Phys. Rev. C} {\bf 68} (2003) 035501.
\bibitem{KOR07}
M. Kortelainen and J. Suhonen, Phys, Rev. C {\bf 75} (2007) 051303(R).
\bibitem{KOR07a}
M. Kortelainen and J. Suhonen, Phys, Rev. C {\bf 76} (2007) 024315.
\bibitem{SIM08}
F. Simkovic {\it et al.}, Phys, Rev. C {\bf 77} (2008) 045503.
\bibitem{BAR09}
J. Barea and F. Iachello, {\it Phys. Rev. C} {\bf 79} (2009) 044301.
\bibitem{RAT10}
P.K. Rath et al., {\it Phys. Rev. C} {\bf 82} (2010) 064310.
\bibitem{ROD10}
T.R. Rodrigues and G.M. Martinez-Pinedo, {\it Phys. Rev. Lett.} {\bf 105} 
(2010) 252503. 
\bibitem{CAU08}
E. Caurier et al., {\it Phys. Rev. Lett.} {\bf 100} (2008) 052503 .
\bibitem{CAU07}
E. Caurier, F. Nowacki and A. Poves, {\it Int. J. Mod. Phys. E} {\bf 16} (2007) 552.
\bibitem{ARNA04}
C. Arnaboldi et al., {\it Nucl. Instrum. Methods A} {\bf 518} (2004) 775.
\bibitem{ABT04}
I. Abt et al., hep-ex/0404039.
\bibitem{MAJ03}
Majorana Collaboration, nucl-ex/0311013.
\bibitem{AVI10}
F. T. Avignone, {\it Prog. Part. Nucl. Phys.} {\bf 64} (2010) 258.	 
\bibitem{DAN00}
M. Danilov et al., {\it Phys. Lett. B} {\bf 480} (2000) 12.
\bibitem{BAR02}
A. S. Barabash, {\it Czech. J. Phys.} {\bf 52} (2002) 575.
\bibitem{BAR04a}
A.S. Barabash, {\it Phys. At. Nucl.} {\bf 67} (2004) 1984.
\bibitem{SAA09}
R. Saakyan, {\it J. Phys. Conf. Ser.} {\bf 179} (2009) 012006.
\bibitem{NAK10a}
K. Nakamura, Report on International Conference "Neutrino'2010", Athens, Greece, 13-19 June,2010.
\bibitem{KRA10}
C. Kraus and S.J.M. Peeters, {\it Prog. Part. Nucl. Phys.} {\bf 64} (2010) 273.
\bibitem{BAR11a} 
A.S. Barabash, {\it Phys. Part. Nucl.} {\bf 42} (2011) 613.
\bibitem{MEI06}
D.-M. Mei and A. Hime, {\it Phys. Rev. D} {\bf 73} (2006) 053004. 
\bibitem{ZDE04}
Y.G. Zdesenko et al., {\it J. Phys. G} {\bf 30} (2004) 971.
\bibitem{ZUB11}
N.F. Barros and K. Zuber, hep-ph/1103.5757.
\bibitem{KLI04}
A. Klimenko, hep-ph/0407156.



\end{thebibliography}

\IfFileExists{\jobname.bbl}{}
 {\typeout{}
  \typeout{******************************************}
  \typeout{** Please run "bibtex \jobname" to optain}
  \typeout{** the bibliography and then re-run LaTeX}
  \typeout{** twice to fix the references!}
  \typeout{******************************************}
  \typeout{}
 }

\end{document}